\documentclass[%
 reprint,
nofootinbib,
 amsmath,amssymb,
 aps,
]{revtex4-2}

\usepackage{graphicx}
\usepackage{dcolumn}
\usepackage{bm}
\usepackage{color}
\usepackage{multirow}
\usepackage{makecell}

\usepackage[dvipsnames]{xcolor}

\begin{document}

\preprint{APS/123-QED}

\title{Safety of Quark/Gluon Jet Classification}

\author{Alexis Romero}
\affiliation{Department of Physics and Astronomy, University of California, Irvine, CA 92627}
\author{Julian Collado}
\affiliation{Department of Computer Science, University of California, Irvine, CA 92627}
\author{Daniel Whiteson}
\affiliation{Department of Physics and Astronomy, University of California, Irvine, CA 92627}
\author{Michael Fenton}
\affiliation{Department of Physics and Astronomy, University of California, Irvine, CA 92627}
\author{Pierre Baldi}
\affiliation{Department of Computer Science, University of California, Irvine, CA 92627}

\date{\today}

\begin{abstract}
The classification of jets as quark- versus gluon-initiated is an important yet challenging task in the analysis of data from high-energy particle collisions and in the search for physics beyond the Standard Model. The recent integration of deep neural networks operating on low-level detector information has resulted in significant improvements in the classification power of quark/gluon jet tagging models. However, the improved power of such models trained on simulated samples has come at the cost of reduced interpretability, raising concerns about their reliability. We elucidate the physics behind quark/gluon jet classification decisions by comparing the performance of networks with and without constraints of infrared and collinear safety, and identify the nature of the unsafe information by revealing the energy and angular dependence of the learned models.  This in turn allows us to approximate the performance of the low-level networks (by 99\% or higher) using equivalent sets of interpretable high-level observables, which can be used to probe the fidelity of the simulated samples and define systematic uncertainties. 
\end{abstract}

\maketitle


\section{\label{sec:intro}Introduction}

Data from experiments at the Large Hadron Collider (LHC) probe some of the most fundamental questions in modern science, such as the nature of dark matter, the potential unification of forces, and the matter/anti-matter imbalance in the observable universe.  In recent years, the integration of machine learning into data analyses has catalyzed scientific advances in particle physics, as the machine-learned models are able to handle data of greater complexity and higher dimensionality than previously feasible~\cite{Baldi_2016, Guest_2016, Guest_2018, Larkoski_2020}.  However, the improved power of such models often comes at the cost of reduced interpretability.  As most models are trained on simulated samples using supervised learning techniques, physicists are rightly concerned that models may base their classification or regression decisions on portions of the feature space which are poorly described by simulations, or where the modeling is theoretically uncertain.  For this reason, it is important that physicists be able to understand the nature of the information being used, providing confidence in the network decisions, and allowing the assessment of systematic uncertainties. 

One important application of machine learning is the classification of quark versus gluon jets~\cite{PhysRevD.44.2025, Gallicchio_2011, Gallicchio_2013, Aad_2014, Gras_2017, Kasieczka_2019, Andrews_2020}.  While significant efforts have been made to improve classification performance, less attention~\cite{Choi:2018dag} has been paid to understanding the Quantum Chromodynamics (QCD) nature of the information being used.  Two main classes of QCD effects can be considered: perturbative and non-perturbative.  Although there is no universal and gauge-invariant way to distinguish between the two, physicists have relied on principles such as infrared and collinear (IRC) safety to identify perturbative effects, as IRC-safety ensures that observables are invariant to soft emissions and arbitrary collinear parton splittings.  Studies in simulated quark/gluon dijet samples indicate that the task of quark/gluon tagging should be dependent only on IRC-safe observables, as the likelihood for quark vs. gluon classification is IRC-safe~\cite{Larkoski:2019nwj}.  Nevertheless, networks trained on high-level information of quark/gluon jet samples simulated with hadronization effects often outperform those trained on high-level information of samples simulated before hadronization (``at parton level"), indicating that quark/gluon tagging is highly sensitive to both perturbative and non-perturbative effects~\cite{Larkoski:2019nwj, Gras_2017}.  Given that the current theoretical understanding of non-perturbative effects like hadronization is limited, as well as a lack of a current unambiguous definition of quark and gluon jets at the hadron level,
variations in the modeling of non-perturbative effects with different parton-shower event generators can occur, thus hindering our abilities to calculate the systematic uncertainties of the low-level models.

Here we focus directly on the question of the nature of the QCD information used by networks that learn to classify quark and gluon jets from low-level calorimeter information.  Our strategy is to identify the source and importance of the perturbative and non-perturbative effects by comparing the performance of networks whose internal structures enforce IRC-safety to those which are unconstrained, on benchmark problems with and without hadronization effects, and as a function of jet energy.  A drop in performance for the networks that enforce IRC-safety relative to the unconstrained networks is attributed to IRC-unsafe information.  We then attempt to more specifically identify the nature of the IRC-unsafe information by employing networks whose internal structure enforces prescribed IRC-unsafe dependencies on transverse momentum and angular distance metrics.  This strategy allows us to reveal the nature of the IRC-unsafe information by narrowing down its energy and angular dependence, enabling us to map it to well-known IRC-unsafe jet variables, such as tower multiplicity, IRC-unsafe generalized angularities~\cite{Larkoski_2014}, and energy-flow polynomials (EFPs)~\cite{Komiske_2018}.  Capturing the information used by low-level networks into high-level observables is vital for future applications of machine learning classifiers~\cite{Faucett:2020vbu}, as it enables physicists to both understand the nature of the information and improve confidence in the network's decisions by allowing for intentional inclusion or exclusion of such information.

The rest of this paper is organized as follows. Section \ref{sec:StrategyAndMethods} lays out the strategy used to reveal the reliance of the low-level networks on IRC-safe and IRC-unsafe information, and for capturing this information into high-level observables.  Section \ref{sec:Data} describes the data generation settings.  Results and discussion are presented in Section \ref{sec:Results}, and conclusions in Section \ref{sec:Conclusions}.

\section{\label{sec:StrategyAndMethods} Strategy and Methods}

Two key elements for understanding the nature of the information used by machine learning classifiers for jet tagging are the inherent constraints of the classifier and the jet representation. 

In recent studies, many jet representations have been considered;  popular choices include: jet images~\cite{de_Oliveira_2016, Baldi_2016, Komiske_2017}, unordered sets of constituents~\cite{Komiske_2019, Qu_2020}, and ordered sets of constituents~\cite{Guest_2016, Louppe_2019, Cheng_2018, egan2017long}.  In order to compare each of the learning strategies on equal footing and to maintain the maximal amount of information, we choose to represent the jets as unordered sets of calorimeter towers.  The towers in the sets are characterized by their three-momenta -- ($p_\textrm{T}, \eta, \phi$) -- and are centered with respect to the $E-$scheme jet axis.  Only towers within a radial distance of $R =\sqrt{\Delta\phi^2+\Delta\eta^2} < 0.4$ from the jet axis are kept. 

The first step in our strategy focuses on assessing the importance of IRC-safe information calculated from the unordered sets of calorimeter towers.  First, we estimate an effective upper limit on the performance of quark/gluon classifiers by employing Particle-Flow Networks (PFNs)~\cite{Komiske_2019}.  PFNs have consistently achieved top performances for quark/gluon jet classification ~\cite{Komiske_2019, Kasieczka_2019, Qu_2020, bogatskiy2020lorentz}, so their performance is taken to be the benchmark to which all other networks in this paper are compared.  Next, we consider Energy-Flow Networks (EFNs)~\cite{Komiske_2019} which, like PFNs,  treat jets as unordered sets of constituents, but use an architecture which constrains internal functions to forms which enforce IRC-safety.  We assess the importance of IRC-safe information by comparing the difference in performance between the PFNs and the EFNs.  A reduced performance by the EFNs relative to the PFNs would suggest that the PFNs are relying on IRC-unsafe information.   

The second step focuses on exploring the nature of the IRC-unsafe information.   We begin by introducing EFN$[\kappa]$s, a generalization of EFNs whose architecture constrains the models to use internal functions with a given energy weighting exponent, $\kappa$.  Variation of the network performance with $\kappa$ reveals the nature of the functional forms which capture the IRC-unsafe information.  We then attempt to map this information onto families of IRC-unsafe observables, which can be concatenated to IRC-safe observables to match the performance of the PFNs.  Two families of IRC-safe observables are used: N-subjettiness variables in combination with jet mass, and IRC-safe EFPs.  Similarly, two families of IRC-unsafe observables are used: IRC-unsafe generalized angularities and IRC-unsafe EFPs, both in combination with tower multiplicity. 

The networks employed in our analysis of the information used during quark/gluon jet classification are detailed below. 

\subsection{Particle-Flow Networks}

The power of PFNs relies on their ability to learn virtually any symmetric function of the towers.  Their mathematical structure is naturally invariant under permutation of the input ordering, as it is built on a summation over the towers.  PFNs can be mathematically summarized as
\begin{equation}
    \textrm{PFN} : F \left( \sum_{i \in \text{jet}}  \Phi(p_{\textrm{T}i}, \eta_i, \phi_i) \right),
    \label{eq:PFN}
\end{equation}
where $\Phi$ represents the per-tower latent space and $F$ the event-level latent space.  The transverse momentum, pseudorapidity, and azimuthal angle of tower $i$ are respectively given by $p_{\textrm{T}i}$, $\eta_i$, and $\phi_i$, and this notation is used throughout this paper when indexing over the towers in a jet.  

We place no constraints on the nature of the latent spaces, giving the network great flexibility.  For this reason, they are a useful probe of the effective upper limit of the performance in the classification tasks when minimal constraints are applied to the nature of the learning method\footnote{We verified in several cases that similar performance is achieved by convolutional neural networks operating on jet images, but due to the significantly increased computational cost and number of parameters, we set them aside as probes of the upper bound in favor of PFNs.}.

\subsection{Energy-Flow Networks}

Similar in structure to PFNs, EFNs are constructed such that the event-level latent space learns functions which have a linear energy factor.  Mathematically, EFNs can be summarized as 
\begin{equation}
    \textrm{EFN} : F \left( \sum_{i \in \text{jet}}  p_{\textrm{T}i} \Phi(\eta_i, \phi_i) \right),
    \label{eq:EFN}
\end{equation}
where unlike in PFNs, $\Phi \left( \eta_i, \phi_i \right)$ is a function only of the angles, weighted by a linear term in transverse momentum, $p_{\textrm{T}i}$, as required for IRC-safety. 

\subsection{IRC-unsafe Energy-Flow Networks}

In anticipation of the importance of IRC-unsafe information, we introduce a generalization of EFNs which includes non-linear energy weighting exponents.  Mathematically, Eq. \ref{eq:EFN} is modified to be non-linear in $p_{\textrm{T}i}$ as
\begin{equation}
    \textrm{EFN}[\kappa] : F \left( \sum_{i \in \text{jet}}  p_{\textrm{T}i}^{\kappa} \Phi(\eta_i, \phi_i) \right),
    \label{eq:EFNkappa}
\end{equation}
where EFN$[\kappa]$ refers to the modified form of an EFN with an energy factor of degree $\kappa$, which is referred to as the \textit{energy weighting exponent} in this paper\footnote{Two equivalent approaches can be used to implement EFN$[\kappa]$: (1) elevating the tower's transverse momentum from $p_{\textrm{T}}$ to $p_{\textrm{T}}^{\kappa}$, and then then passing them as input to the EFN, and (2) directly modifying the architecture of the EFN to use $p_{\textrm{T}}^{\kappa}$ as the weighting parameter of $\Phi(\eta_i, \phi_i)$ in Eq. \ref{eq:EFN}.}.  This modification allows us isolate the critical values of $\kappa \neq 1$ that capture most of the IRC-unsafe information needed for the quark/gluon classification task.  Note that EFN$[\kappa]$ can also be used to explore IR-safe energy weighting factors by setting $\kappa > 1$.

\subsection{\label{sec:JetsAsHlFeatures} High-level Observables}

The literature on strategies for quark/gluon classification using high-level observables is quite mature, providing many families of high-level observables which reduce the large dimensionality of the input space into one-dimensional observables that retain useful information for classification.  These observables have the significant advantages that they are physically interpretable, compact, and allow for reasonable assessment of systematic uncertainties due to, for example, mismodeling in simulation. The disadvantage is that they are limited to those ideas conceived of by human physicists.  In our study, the observables are calculated directly from the calorimeter towers, and are paired with fully connected Dense Neural Networks (DNNs) or with Linear Discriminant Analysis (LDA) classification models in the cases where the observables are linearly separable, such as EFPs.

We consider the following IRC-safe observables that are traditionally used in quark/gluon studies:

\begin{itemize}
    \item N-subjettiness variables~\cite{Thaler_2011, Thaler_2012}, which provide a measure of the degree to which the radiation within a jet is aligned along $N$ candidate subjet axes, and are defined as
    \begin{equation}
        \tau_N[\beta] = \sum\limits_{i \in \textrm{jet}} p_{\textrm{T}i} \textrm{min} \{ R_{i, 1}^{\beta}, R_{i, 2}^{\beta}, \ldots , R_{i, N}^{\beta} \}, 
        \label{eq:Nsub}
    \end{equation}
    where $R_{i, J}$ is the angular distance between subjet axis $J$ $(J \leq N)$ and tower $i$. The parameter $\beta$ $(\beta>0)$ is referred to as the \textit{angular weighting exponent} in this paper.  Following~\cite{Datta_2017}, we compute the first 18 N-subjettiness observables with respect to the $k_\textrm{T}$ axis:
    \begin{flalign*}
        \{ & \tau_1{[\beta=\frac{1}{2}]}, \tau_1{[\beta=1]}, \tau_1{[\beta=2]}, \ldots, \\
        & \tau_6{[\beta=\frac{1}{2}]}, \tau_6{[\beta=1]}, \tau_6{[\beta=2]} \}.
    \end{flalign*}
\end{itemize}

\begin{itemize}
    \item Jet mass ($m_{\textrm{jet}}$), which has been found to be a powerful quark/gluon jet discriminant~\cite{Gallicchio_2011}.
\end{itemize}

\begin{itemize}
    \item IRC-safe Energy-Flow Polynomials~\cite{Komiske_2018}, which are sets of non-isomorphic multigraphs that linearly span the space of IRC-safe observables. For a multigraph $G$ with $V$ vertices and edges $(k, l) \in G$, the corresponding EFP observable is defined as
    \begin{equation}
        \textrm{EFP}[\beta] = \sum\limits_{i_1 \in \textrm{jet}} \cdots \sum\limits_{i_V \in \textrm{jet}} p_{\textrm{T}i_1} \cdots p_{\textrm{T}i_V} \prod\limits_{(k,l) \in G} R_{i_k, i_l}^{\beta} 
        \label{eq:EFPsafe}
    \end{equation}
    where $R_{i j}$ is the angular distance between particles $i$ and $j$.  Following~\cite{Komiske_2018}, the optimal performance for quark/gluon jet classification using IRC-safe EFPs is achieved with $\beta=\frac{1}{2}$; we employ the same set for our studies, with a maximum number of edges of $d \leq 5$, where $d$ corresponds to the degree of the angular monomial.
\end{itemize}

\begin{figure*}
    \centering
    \includegraphics{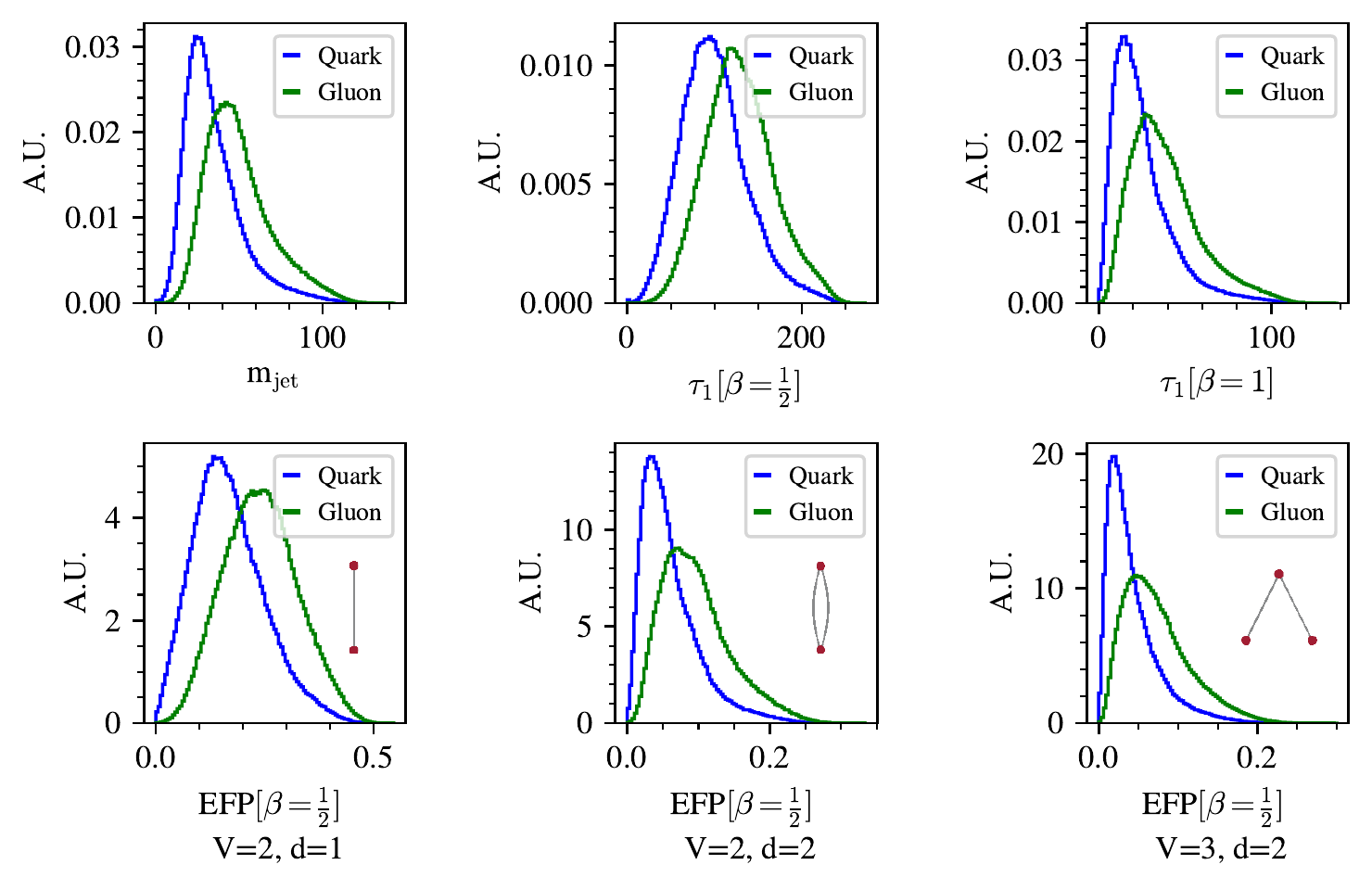}
    \caption{ Distributions of jet mass and select N-subjettiness observables (top), and IRC-safe EFP observables (bottom), for quark- and gluon-initiated jets with $p_{\textrm{T}} \in [500, 550]$ GeV. The shape of the EFP multigraphs is shown for illustrative purposes.}
    \label{fig:DistributionsSafe}
\end{figure*}

Distributions of a selection of N-subjettiness and IRC-safe EFP observables are shown in Fig.~\ref{fig:DistributionsSafe}.

To capture the IRC-unsafe information, we consider the following IRC-unsafe observables:

\begin{itemize}
    \item IRC-unsafe Generalized Angularities~\cite{Larkoski_2014}, which have a simple form that allows for easy interpretation, are defined as 
    \begin{equation}
        \lambda[\kappa, \beta] = \sum\limits_{i \in \text{jet}} p_{\textrm{T}i}^{\kappa} 
        \left( \frac{R_{i,\textrm{jet}}}{R} \right)^{\beta},
        \label{eq:GenAng}
    \end{equation}
    where $R_{i, \text{jet}}$ is the radial distance from tower $i$ to the $k_T$ jet axis, and $R$ is the jet radius.
\end{itemize}

\begin{itemize}
    \item IRC-unsafe Energy-Flow Polynomials~\cite{Komiske_2018}, which have a similar form to the IRC-safe EFPs in Eq. \ref{eq:EFPsafe}, but with a non-linear energy weighting exponent ($\kappa \neq 1$).  Following the notation in Eq. \ref{eq:EFPsafe}, the IRC-unsafe EFPs are defined as
    \begin{equation}
        \textrm{EFP}[\kappa, \beta] = \sum\limits_{i_1 \in \textrm{jet}} \cdots \sum\limits_{i_N \in \textrm{jet}} p_{\textrm{T}i_1}^{\kappa} \cdots p_{\textrm{T}i_N}^{\kappa} \prod\limits_{(k,l) \in G} R_{i_k i_l}^{\beta}.
        \label{eq:EFPunsafe}
    \end{equation}
\end{itemize}

\begin{itemize}
    \item Tower multiplicity $(n_{\mathrm{t}})$, which counts the number of towers in a jet and has also been found to be a powerful quark/gluon jet discriminant~\cite{Gallicchio_2011}.
\end{itemize}

\begin{figure*}
    \centering
    \includegraphics{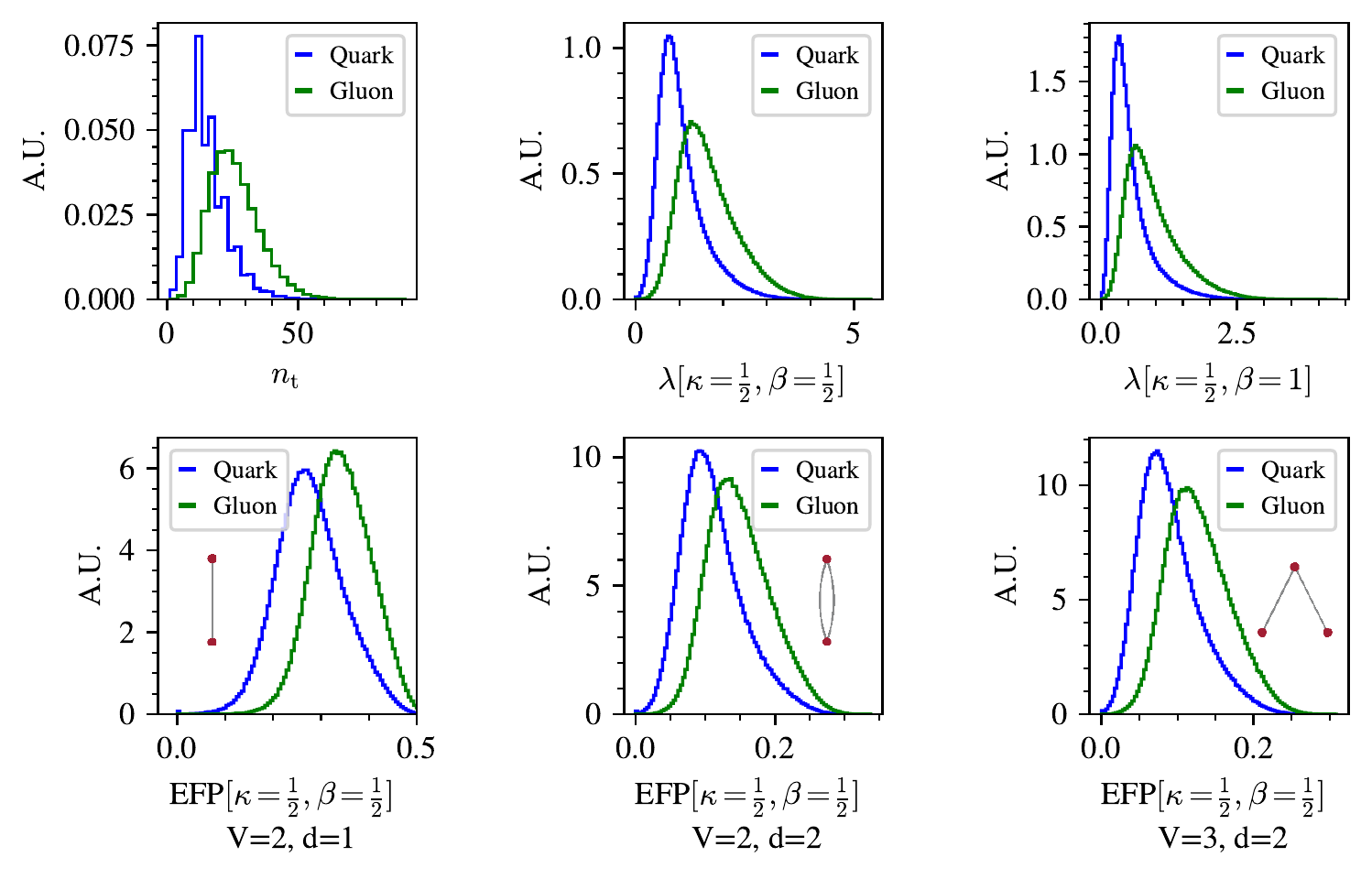}
    \caption{ Distributions of tower multiplicity and select IRC-unsafe Generalized Angularity observables (top), and IRC-unsafe EFP observables (bottom), for quark- and gluon-initiated jets with $p_{\textrm{T}} \in [500, 550]$ GeV. The shape of the EFP multigraphs is shown for illustrative purposes.}
    \label{fig:DistributionsUnsafe}
\end{figure*}

Distributions of a selection of IRC-unsafe generalized angularity and IRC-unsafe EFP observables are shown in Fig.~\ref{fig:DistributionsUnsafe}.

A summary of all the models trained with high-level observables is shown in Table~\ref{tab:ModelSummaries}.

{\renewcommand{\arraystretch}{1.2}
\begin{table*}
\caption{\label{tab:ModelSummaries} Summary of the models trained on IRC-safe and IRC-unsafe jet observables.}
\begin{tabular}{llr}
\hline
\hline
Model Name & Description & \makecell[l]{Number of \\ Observables}\\ 
\hline
DNN[safe] &
DNN trained on N-subjettiness variables and jet mass. & 
19 \\
DNN[safe, $n_{\mathrm{t}}$, $\lambda[\frac{1}{2}, \beta]$] &
\makecell[l]{DNN trained on N-subjettiness variables, jet mass, tower multiplicity, and  \\  
a generalized angularity variable with $\kappa=\frac{1}{2}$ and $\beta \in \{\frac{1}{2}, 1, 2\}$.} & 21 \\ 
LDA[safe] &
LDA trained on IRC-safe EFP variables ($d \leq 5$) with $\beta=\frac{1}{2}$. & 
102 \\
LDA[safe, $n_{\mathrm{t}}$, EFP[$\frac{1}{2}$, $\beta$]] &
\makecell[l]{LDA trained on IRC-safe EFP variables ($d \leq 5$) with $\beta=\frac{1}{2}$, tower multiplcity, \\ 
and IRC-unsafe EFP variables ($d \leq 5$) with $\kappa=\frac{1}{2}$ and $\beta \in \{\frac{1}{2}, 1, 2\}$.} & 
205 \\
\hline \hline
\end{tabular}
\end{table*}
}

\section{Datasets and Training}
\label{sec:Data}
Samples of light quark (\textit{u, d, s}) jets and gluon jets  are generated in dijet events from $pp$ collisions at $\sqrt{s}$=14 TeV.  Collisions and immediate decays are generated with \textsc{Madgraph5} v2.6.5~\cite{Alwall_2014}, while showering and hadronization is simulated with \textsc{Pythia} v8.235~\cite{Sj_strand_2008}.  The light quark-initiated jets come from the parton level hard-processes $pp \rightarrow qq$ and $q\bar{q}$ while the gluon-initiated jets come from $pp \rightarrow gg$.  Mixed quark and gluon states are not generated to minimize ambiguity, as precise theoretical definitions of quark/gluon jet labels are generally elusive~\cite{Gras_2017}, though operational jet flavor definitions~\cite{Metodiev_2018, Komiske_2018_opdef} may be used to classify quark and gluon jets directly in LHC data.  To compare the effects of particle hadronization on quark/gluon jet tagging, the events are generated with and without hadronization effects, respectively corresponding to ``hadron'' and ``parton'' level events, by toggling the {\sc HadronLevel:all} switch in \textsc{Pythia}.  Jets are then passed through the \textsc{Delphes} v3.4.2~\cite{deFavereau:2013fsa} detector simulator, using the standard ATLAS card, to simulate interactions with the detector\footnote{Note that detector simulations such as Delphes~\cite{deFavereau:2013fsa} may introduce low-$p_{\textrm{T}}$ cutoffs, which effectively act as controlled cutoffs for IRC-unsafe observables.}.  No additional $pp$ interactions (pileup) are considered, as many studies have shown effective mitigation techniques to attenuate the effects of pileup ~\cite{Aad_2016, Komiske_pileup_2017, martinez2019pileup}. 

Jets are reconstructed from calorimeter towers with the anti-$k_\textrm{T}$ clustering algorithm~\cite{Cacciari:2008gp}, as implemented in \textsc{FastJet} v3.3.2~\cite{Cacciari:2011ma}, with a distance parameter of $R=0.4$ and disregarding neutrinos.  In each event, only the hardest jet with absolute pseudorapidity $|\eta| < 2.0$ is kept. 

To study energy dependence, three ranges of jet $p_\textrm{T}$ are considered: $200-220$ GeV, $500-550$ GeV, and $1000-1100$ GeV, with the threshold applied to reconstructed jets.  For efficiency of generation, similar thresholds are applied at parton-level, but with a window 20\% broader to avoid distortions.  

For each jet $p_\textrm{T}$ range, 650k quark jets and 650k gluon jets are generated; these are split into datasets with 1M events for training, 200k for testing, and 100k for validation. The sets of unordered towers used as inputs to the low-level networks -- PFN, EFN, and EFN${[\kappa]}$ -- are preprocessed by normalizing the sum of the $p_\textrm{T}$ of the towers in the sets to unity.  The observables used as inputs to the high-level classifiers -- DNN and LDA -- are preprocessed by subtracting the mean and dividing by the standard deviation of the distributions in the training set.  See the Appendix for details on network architectures and training.  The performance of the various classification strategies is compared using the area under the receiver operating curve (AUC) of each network.  The statistical uncertainty on the strategies is measured using boostraping to $\pm$ 0.002 or less, unless otherwise specified.

\section{\label{sec:Results}Results and Discussion}

The PFNs, which provide a loose upper limit, perform well, with classification power increasing with jet $p_\textrm{T}$ as shown in Tab.~\ref{tab:AUCs_hadON}.  The EFNs, which are limited to IRC-safe information, show a small but statistically significant drop in relative performance.  The difference in the EFNs and PFNs internal constraints allows us to conclude that the difference in performance is due to the use of IRC-unsafe information by the PFNs, which grows modestly in importance with jet $p_\textrm{T}$.
 
\begin{figure}
    \centering
    \includegraphics{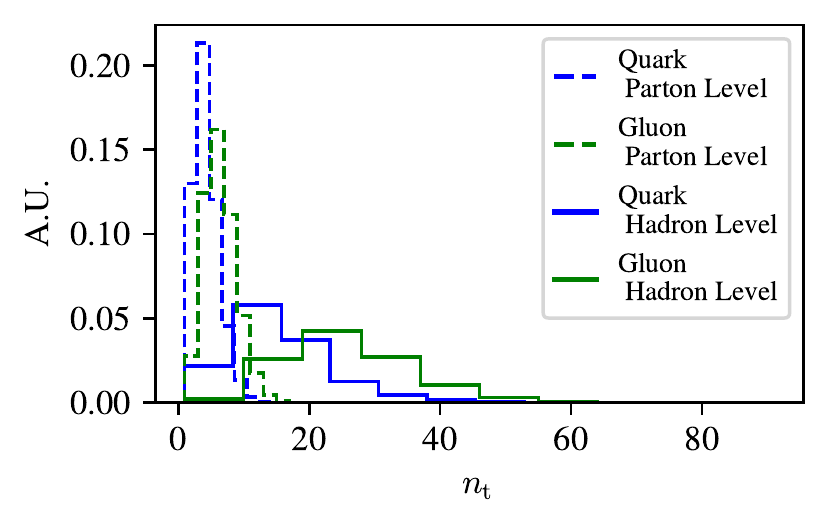}
    \caption{Distributions of tower multiplicity ($n_{\mathrm{t}}$) for quark- and gluon-initiated jets with $p_{\textrm{T}} \in [500, 550]$ GeV, simulated with hadronization effects (solid line) and without hadronization effects (dashed line).}
    \label{fig:nTowers_had_noHad}
\end{figure}

To understand the physical source of the IRC-unsafe information, we train EFN and PFN networks on events at parton and hadron level.  In the parton level events, the PFN-EFN gap vanishes (see Tab.~\ref{tab:AUCs_hadOFF}), confirming the results of Ref.~\cite{Bieringer:2020tnw} and demonstrating the central conclusions of Ref.~\cite{Larkoski:2019nwj};  without hadronization effects only IRC-safe information is needed for quark-gluon tagging. In addition, this comparison reveals that the IRC-unsafe information is introduced in the non-perturbative hadronization process.  The contrast in quark and gluon jets simulated with and without hadronization effects can be seen in observables sensitive to the number of non-perturbative emissions, such as tower multiplicity, as illustrated in Fig. \ref{fig:nTowers_had_noHad}.  The number of towers in quark and gluon jets increases when including hadronization effects, indicating that despite low-$p_{\textrm{T}}$ detector cutoffs, calorimeter towers may be sensitive to non-perturbative emissions, resulting in statistically significant contributions to quark/gluon jet classifiers. 

\begin{table}
\caption{\label{tab:AUCs_hadON} Comparison of the quark-gluon classification performance of EFN and PFN networks, via AUC, on jets with hadronization effects included.}
\begin{ruledtabular}
\begin{tabular}{rccc}
Jet $p_{\textrm{T}}$ Range & EFN & PFN &$\Delta$(PFN-EFN) \\ \hline
200-220 GeV  & 
0.814 $\pm$ 0.001 &
0.828 $\pm$ 0.001 &
0.014 $\pm$ 0.002\\
500-550 GeV  & 
0.819 $\pm$ 0.001 &
0.838 $\pm$ 0.001 &
0.019 $\pm$ 0.002\\
1000-1100 GeV & 
0.827 $\pm$ 0.001 &
0.848 $\pm$ 0.001 &
0.021 $\pm$ 0.002\\
\end{tabular}
\end{ruledtabular}
\end{table}

\begin{table}
\caption{\label{tab:AUCs_hadOFF}Comparison of the quark-gluon classification performance of EFN and PFN networks, via AUC, on jets with no hadronization effects included.}
\begin{ruledtabular}
\begin{tabular}{rccc}
Jet $p_{\textrm{T}}$ Range & EFN & PFN &$\Delta$(PFN-EFN) \\ \hline
200-220 GeV  & 
0.739 $\pm$ 0.001 &
0.737 $\pm$ 0.001 &
-0.002 $\pm$ 0.002\\
500-550 GeV  & 
0.753 $\pm$ 0.001 &
0.750 $\pm$ 0.001 &
-0.003 $\pm$ 0.002\\
1000-1100 GeV & 
0.759 $\pm$ 0.001 &
0.758 $\pm$ 0.001 &
-0.001 $\pm$ 0.002\\
\end{tabular}
\end{ruledtabular}
\end{table}

\begin{table*}
\caption{ Comparison of the quark-gluon classification performance of EFN${[\kappa]}$, classified by $p_{\textrm{T}}$ range and $\kappa \in \{-1, -\frac{1}{2}, -\frac{1}{4}, \frac{1}{4}, \frac{1}{2}, \frac{3}{2}\}$. See Fig.~\ref{fig:KappaSearch} for a visual representation and comparison to EFP and PFN performance.}
\label{tab:AUCs_KappaSearch} 
\begin{ruledtabular}
\begin{tabular}{ccccccc}
Jet $p_{\textrm{T}}$ range & 
\vspace{0.05cm}
\makecell[c]{EFN${[\kappa=-1]}$} &
\makecell[c]{EFN${[\kappa=-\frac{1}{2}]}$} &
\makecell[c]{EFN${[\kappa=-\frac{1}{4}]}$} &
\makecell[c]{EFN${[\kappa=\frac{1}{4}]}$} &
\makecell[c]{EFN${[\kappa=\frac{1}{2}]}$} &
\makecell[c]{EFN${[\kappa=\frac{3}{2}]}$} \\ \hline
200-220 GeV  & 
0.785 $\pm$ 0.001 &
0.788 $\pm$ 0.001 &  
0.794 $\pm$ 0.001 &
0.817 $\pm$ 0.001 &
0.821 $\pm$ 0.001 &
0.815 $\pm$ 0.001 \\
500-550 GeV  & 
0.796 $\pm$ 0.001 &
0.802 $\pm$ 0.001 &
0.811 $\pm$ 0.001 &
0.830 $\pm$ 0.001 &
0.831 $\pm$ 0.001 &
0.811 $\pm$ 0.001 \\
1000-1100 GeV & 
0.801 $\pm$ 0.001 &
0.809 $\pm$ 0.001 &
0.822 $\pm$ 0.002 &
0.842 $\pm$ 0.001 &
0.841 $\pm$ 0.001 &
0.812 $\pm$ 0.001 \\
\end{tabular}
\end{ruledtabular}
\end{table*}

\begin{figure*}
    \includegraphics{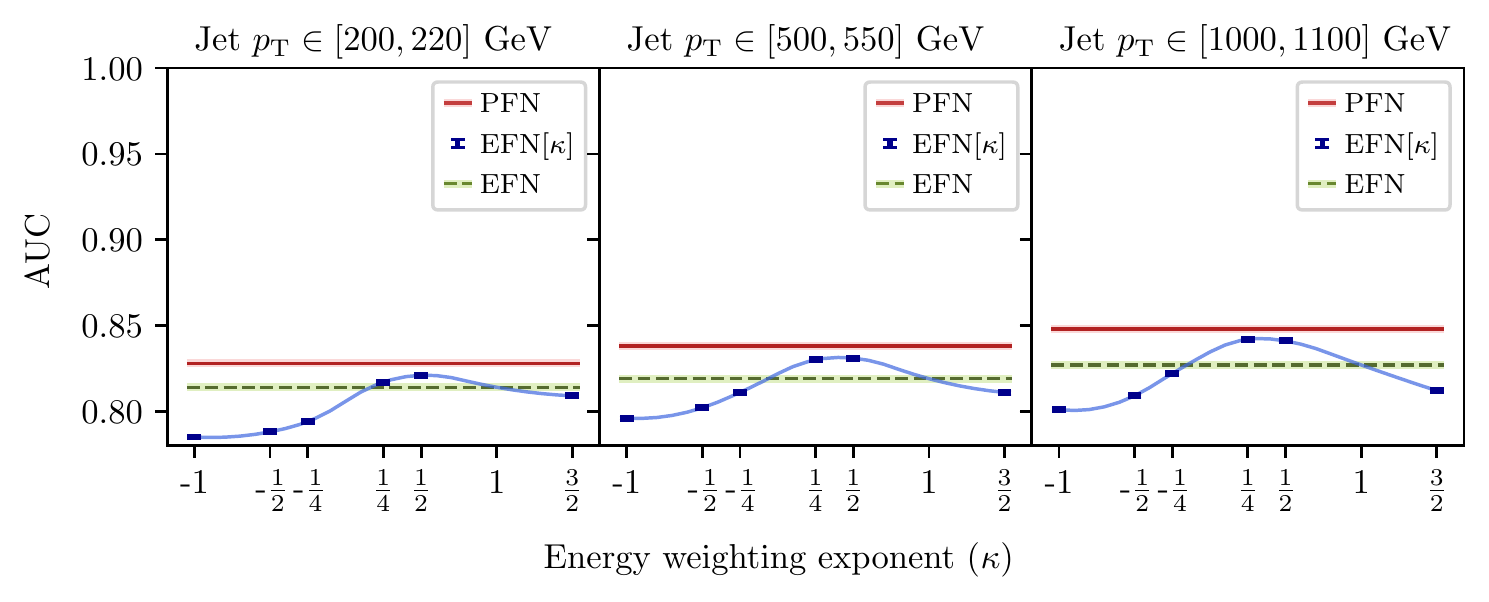}
    \caption{ Comparison of the quark-gluon classification performance, measured by AUC, of the EFN${[\kappa]}$s, for several choices of the energy weighting exponent $\kappa \in \{-1, -\frac{1}{2}, -\frac{1}{4}, \frac{1}{4}, \frac{1}{2}, \frac{3}{2}\}$, which reveals the exponent necessary to exceed the performance of the IRC-safe EFN (dashed green) and approach the performance of the IRC-unsafe PFN (solid red).  Solid blue lines are a quadratic interpolation of the measurements of the EFN${[\kappa]}$ performance at each value of $\kappa$, also given in Table~\ref{tab:AUCs_KappaSearch}. Three panels show the performance in the distinct jet $p_{\textrm{T}}$ ranges.
    }
    \label{fig:KappaSearch}
\end{figure*}

PFNs are very flexible networks, allowing a vast space of possible functions. To understand how the PFNs capture the IRC-unsafe information, we seek to narrow the scope of possible functional forms.  
First, we attempt to narrow down the energy weighting exponents of the necessary IRC-unsafe information by comparing the performance of EFN${[\kappa]}$s for the range of values $\kappa \in \{-1, -\frac{1}{2}, -\frac{1}{4}, \frac{1}{4}, \frac{1}{2}, \frac{3}{2}\}$.  The selected range covers both softer and harder radiation, as large values of ${\kappa}$ accentuate harder hadrons while small values of ${\kappa}$ accentuate softer hadrons.  

As shown in Tab.~\ref{tab:AUCs_KappaSearch} and Fig.~\ref{fig:KappaSearch},  EFN${[\kappa]}$ performs well for energy weighting exponents close to zero, with the best performing values between $\frac{1}{4} \leq \kappa \leq \frac{1}{2}$.  This indicates that the IRC-unsafe information in the PFN-EFN gap is mainly due to soft radiation, and can potentially be captured by observables with small energy weighting exponents.  In addition, we note how soft radiation becomes more relevant with higher jet $p_{\textrm{T}}$, as the EFN${[\kappa=\frac{1}{4}]}$ and the EFN${[\kappa=\frac{1}{2}]}$ increasingly outperform the EFN as energy increases.  For simplicity, we take $\kappa=\frac{1}{2}$ to be the critical energy weighting exponent as it consistently outperforms or effectively matches the other $\kappa$ values.  

Having isolated the critical energy weighting exponent which captures the IRC-unsafe information, the next step is to identify the critical angular weighting exponent, $\beta$.  However, unlike the energy weighting, the EFN structure does not allow us to easily constrain the critical angular weighting values.  Instead, we search for a set of observables with specific angular weighting exponents which can be combined with IRC-safe observables to approximate the PFN performance.  We consider IRC-unsafe observables with energy weighting exponent $\kappa=\frac{1}{2}$ and angular weighting exponent $\beta \in \{\frac{1}{2}, 1, 2\}$, to cover narrow- and wide-angle radiation.  A summary of the high-level models and the corresponding IRC-safe and IRC-unsafe observables used in the search is shown in Table \ref{tab:ModelSummaries}.  

\begin{figure*}[!ht]
    \includegraphics{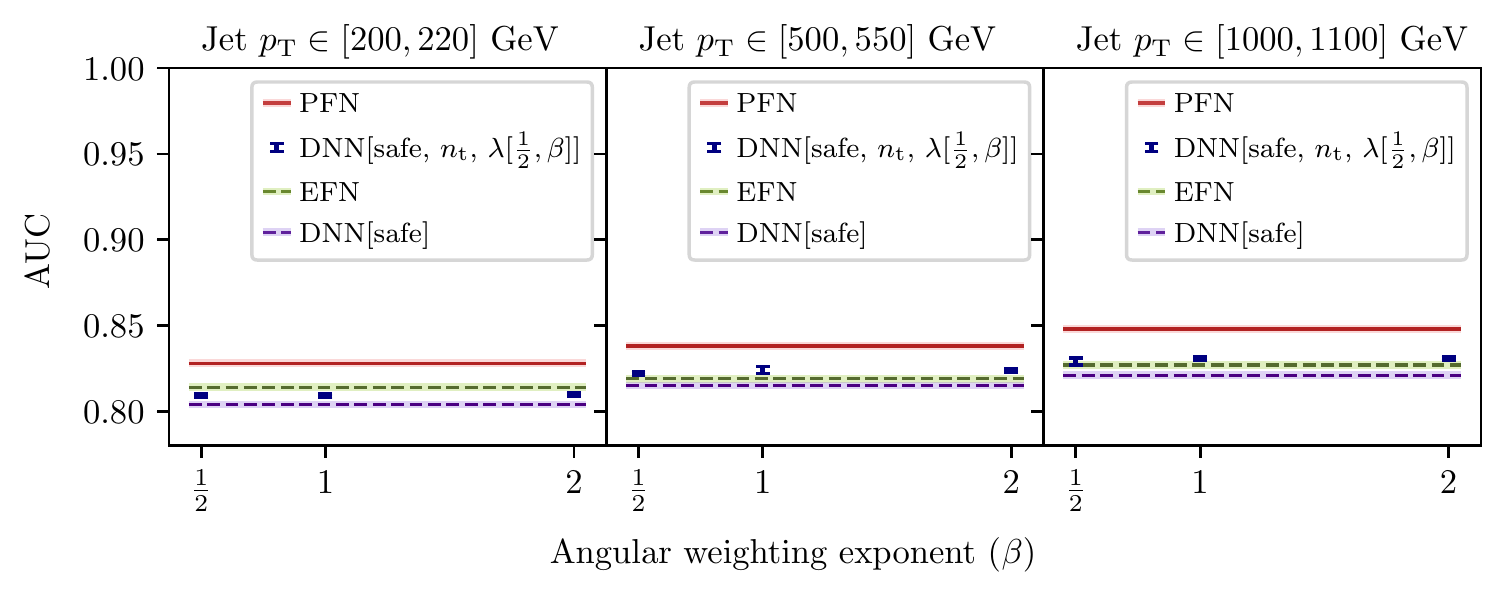}
    \caption{AUCs of the DNN models trained on IRC-safe N-subjettiness and IRC-unsafe generalized angularity observables with $\kappa=\frac{1}{2}$ and $\beta \in \{\frac{1}{2}, 1, 2\}$, in combination with the tower multiplicity.}
    \label{fig:BetaSearch_traditional}
\end{figure*}

The results for the traditional features (N-subjettiness, jet mass, tower multiplicity, and IRC-unsafe generalized angularities) are shown in Table \ref{tab:AUCs_BetaSearch_traditional} and illustrated in Figure \ref{fig:BetaSearch_traditional}.  These traditional observables fail to capture sufficient IRC-safe and IRC-unsafe information to match the PFN in all energy ranges.  

\begin{figure*}[!ht]
    \includegraphics{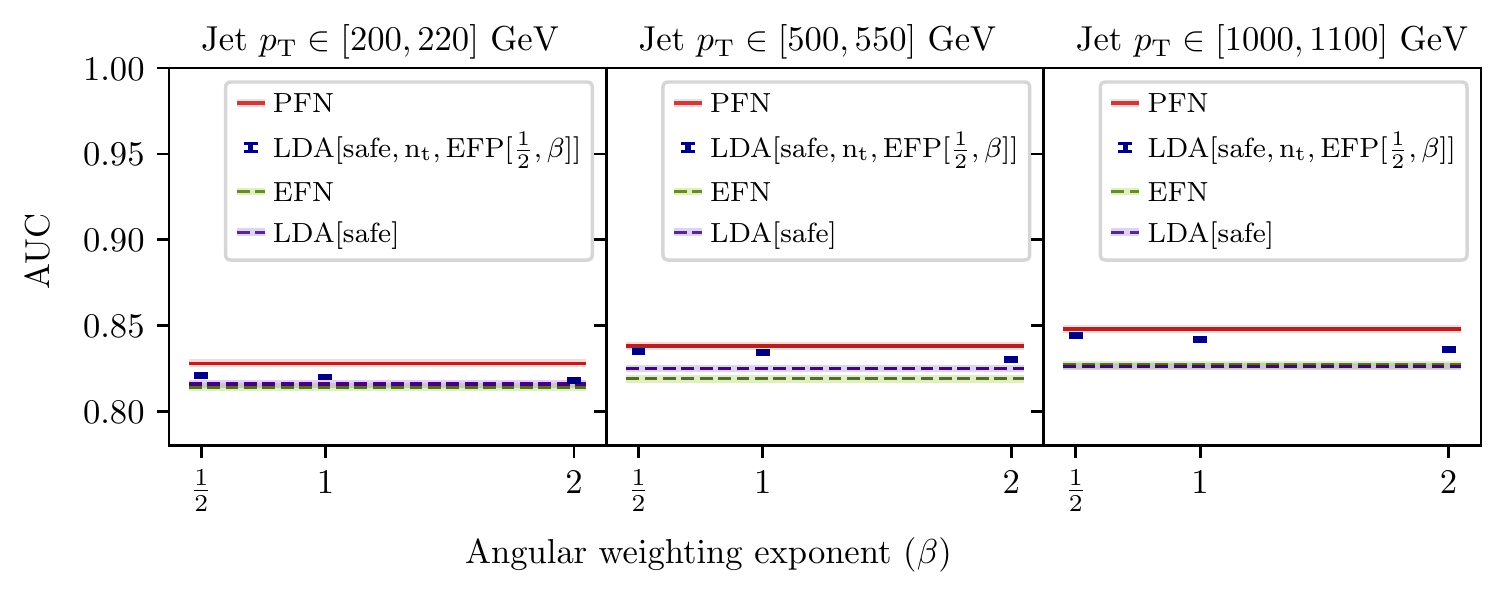}
    \caption{AUCs of the LDA models trained on IRC-safe EFP and IRC-unsafe EFP observables with $\kappa=\frac{1}{2}$ and $\beta \in \{\frac{1}{2}, 1, 2\}$, in combination with the tower multiplicity.}
    \label{fig:BetaSearch_EFPs}
\end{figure*}

The results for the LDA models using EFPs and tower multiplicity are shown in Table \ref{tab:AUCs_BetaSearch_EFPs} and illustrated in Figure \ref{fig:BetaSearch_EFPs}.  In contrast to the traditional features, IRC-safe EFPs largely capture the IRC-safe information used by the EFNs; in addition, there is a boost in performance when combining them with IRC-unsafe EFPs with small angular weighting exponents such as $\beta=\frac{1}{2}$, nearly matching the PFN performances\footnote{LDA models trained on IRC-safe and IRC-unsafe EFPs with $d<=6$ and $d<=7$ are also considered, in each case providing a marginal improvement in AUC, at the cost of significantly more EFP variables. LDA models with EFPs with $d<=5$ are thus chosen in this paper as they result in a good approximation of the PFN performances while having a manageable size of EFP variables.}.  The boost in performance provided by the IRC-unsafe observables increases with energy range, which is consistent with the results illustrated in Fig. \ref{fig:KappaSearch}, corroborating the importance of IRC-unsafe observables for jets with higher $p_\textrm{T}$.  Although less compact than the traditional observables, EFPs are more effective at capturing the necessary information for quark/gluon classification. 

\begin{table*}[!ht]
\caption{\label{tab:AUCs_BetaSearch_traditional} AUCs of the DNN models trained on IRC-safe jet mass N-subjettiness variables, and their combinations with IRC-unsafe generalized angularities with $\kappa=\frac{1}{2}$ and $\beta \in \{\frac{1}{2}, 1, 2\}$.}
\begin{ruledtabular}
\begin{tabular}{ccccc}
Jet $p_{\textrm{T}}$ range & 
DNN[safe] & 
DNN[safe, $n_{\mathrm{t}}$, $\lambda[\frac{1}{2}, \frac{1}{2}]$] &
DNN[safe, $n_{\mathrm{t}}$, $\lambda[\frac{1}{2}, 1]$] &
DNN[safe, $n_{\mathrm{t}}$, $\lambda[\frac{1}{2}, 2]$] \\ \hline 
200-220 GeV  & 
0.804 $\pm$ 0.001 &
0.809 $\pm$ 0.001 &
0.809 $\pm$ 0.001 &
0.810 $\pm$ 0.001 \\
500-550 GeV  & 
0.815 $\pm$ 0.001 &
0.822 $\pm$ 0.001 &
0.824 $\pm$ 0.002 &
0.824 $\pm$ 0.001 \\
1000-1100 GeV & 
0.822 $\pm$ 0.002 &
0.829 $\pm$ 0.002 &
0.831 $\pm$ 0.001 &
0.831 $\pm$ 0.001 \\
\end{tabular}
\end{ruledtabular}
\end{table*}

\begin{table*}[!ht]
\caption{\label{tab:AUCs_BetaSearch_EFPs} AUCs of the LDA models trained on IRC-safe EFPs, and their combinations with tower multiplicity and IRC-unsafe EFPs with $\kappa=\frac{1}{2}$ and $\beta \in \{\frac{1}{2}, 1, 2\}$, with $d \leq 5$ edges.}
\begin{ruledtabular}
\begin{tabular}{ccccc}
Jet $p_{\textrm{T}}$ range & 
LDA[safe] &
LDA[safe, $n_{\mathrm{t}}$, EFP[$\frac{1}{2}, \frac{1}{2}$]] & 
LDA[safe, $n_{\mathrm{t}}$, EFP[$\frac{1}{2}, 1$]] &  
LDA[safe, $n_{\mathrm{t}}$, EFP[$\frac{1}{2}, 2$]] \\ \hline
200-220 GeV  & 
0.816 $\pm$ 0.001 &
0.821 $\pm$ 0.001 &
0.820 $\pm$ 0.001 &
0.818 $\pm$ 0.001 \\
500-550 GeV  & 
0.825 $\pm$ 0.001 &
0.835 $\pm$ 0.001 &
0.834 $\pm$ 0.001 &
0.830 $\pm$ 0.001 \\
1000-1100 GeV & 
0.826 $\pm$ 0.001 &
0.844 $\pm$ 0.001 &
0.842 $\pm$ 0.001 &
0.836 $\pm$ 0.001 \\
\end{tabular}
\end{ruledtabular}
\end{table*}

\section{\label{sec:Conclusions}Conclusions}

In this work, we have confirmed that state-of-the-art machine learning models for quark/gluon jet classification are sensitive to perturbative and non-perturbative effects, the latter rooted in the hadronization process.  Moreover, we have shown that the reliance of the networks on the non-perturbative IRC-unsafe information grows with jet $p_\textrm{T}$.  Although the IRC-unsafe observable space is in principle infinite, its energy and angular dependence can be narrowed down by utilising the strategies introduced in this paper. 

By comparing the performance of networks whose architecture constrains the models to learn functions with prescribed energy weighting forms, EFN${[\kappa]}$s, we have found that most of the IRC-unsafe information can be captured by observables with small energy weighting exponents ($ \frac{1}{4} \leq \kappa \leq \frac{1}{2}$).  Similarly, we have performed a grid search of high-level observables with narrow categories of angular weighting factors to delimit the angular dependence of the IRC-unsafe information. The results show that most of this information can be captured by small angular weighting factors ($\beta=\frac{1}{2}$). This indicates that, as expected, IRC-unsafe information relevant for quark/gluon jet classification is due to soft, narrow-angle radiation. 

Understanding the nature of the information used by deep neural networks trained for the classification of quark- vs. gluon-initiated jets, and mapping it into physics interpretable and compact jet observables, is an extremely powerful tool that could be used for analyses searching for signals beyond the Standard Model.  The strategy presented in this paper allows for the interpretation of the information learned by PFNs in terms of high-level physics observables which provide a sense of the nature of the machine-learned information.  This information was found to be both IRC-safe and IRC-unsafe, corresponding to perturbative and non-perturbative hadronization effects.  The strategy proposed in this paper allows physicists to control and assess the systematic uncertainties of the networks by confidently including or excluding information from the learning process.  In addition, this strategy can easily be extended to other analyses where having robust and interpretable observables that match the performances of deep neural networks would be a powerful tool.

\begin{acknowledgments}
We wish to thank Andrew Larkoski, Ian Mount, Benjamin Nachman, Joakim Olsson, Tilman Plehn, and Jesse Thaler for their valuable feedback and insightful discussions. We also thank Wenjie Huang for his work on the initial stages on this paper. This material is based upon work supported by the National Science Foundation under grant number 1633631. DW and MF are supported by the DOE Office of Science. The work of JC and PB in part supported by grants NSF 1839429 to PB.
\end{acknowledgments}

\appendix
\section{Neural Network Hyperparameters and Architecture}

Common properties across all networks include \textsc{}{ReLu}~\cite{inproceedings} activation functions for all hidden layers, and a sigmoidal output unit at the end to classify between quark and gluon jets.  The low-level networks are trained using stochastic gradient descent and initialized using the He uniform weights~\cite{He_init}.  The DNNs are trained using the Adam optimizer~\cite{Adam2014}, and initialized using Glorot uniform weights~\cite{glorotUniform}.  In all cases, the models are optimized to minimize the relative  entropy  between  the  targets and the outputs across all training examples.  All models are trained for up to 100 epochs with early stopping. The number of layers, nodes in each layer, dropout~\cite{Dropout} rate, and the learning rate are selected using Bayesian Optimization from the Sherpa~\cite{hertel2020sherpa} optimization library.  The ranges for the search of hyperparameters are shown in tables \ref{tab:hyperparam_range_efn_pfn} and \ref{tab:hyperparam_range_hl}.  All models are trained using Keras~\cite{chollet2015keras} from the Tensorflow package~\cite{tensorflow2015}, with batch sizes of 128.

\begin{table}[h]
    \centering
        \caption{Hyperparameter ranges for bayesian optimization of PFN, EFN, and EFN$[\kappa]$ networks.}
            \label{tab:hyperparam_range_efn_pfn}
    \begin{tabular}{c|c}
    \hline\hline
        Parameter &  Range\\
        \hline
          Num. of layers in the per-tower module $\Phi$ & [1, 5] \\
          Size of the layers in the per-tower module $\Phi$ & [32, 196]\\
          Num. of layers in the backend module $F$ & [1, 5] \\
          Size of the layers in the backend module $F$ & [32, 196]  \\
          Learning rate & [0.0001, 0.01] \\
          Learning rate decay & [0.000001, 0.001] \\
          Dropout rate for the layers in module $F$ & [0.0, 0.5] \\
             \hline\hline
    \end{tabular}
\end{table}

\begin{table}[h]
    \centering
        \caption{Hyperparameter ranges for bayesian optimization of DNN networks.}
            \label{tab:hyperparam_range_hl}
    \begin{tabular}{c|c}
    \hline\hline
        Parameter &  Range\\
        \hline
          Num. of layers & [1, 10] \\
          Num. of units & [2, 196]  \\
          Learning rate & [0.0001, 0.01] \\
          Dropout & [0.0, 0.5] \\
             \hline\hline
    \end{tabular}
\end{table}

\appendix

\bibliography{qg_main,baldi}

\end{document}